\documentclass[12pt,thmsa]{article}
\usepackage{sw20aip}

\input tcilatex
\QQQ{Language}{
American English
}

\begin{document}

\author{V. A. Ivanov$^1$, Z. V. Popovic$^{^{\prime }\text{ }2}$, \and O. P. Khuong$%
^3 $, V. V. Moshchalkov \\
\\
Laboratorium voor Vaste-Stoffysica en Magnetisme,\\
Katholieke Universiteit Leuven, \\
Celestijnenlaan 200D, B-3001 Leuven, BELGIUM}
\title{EFFECTIVE DIMENSIONALITY AND BAND STRUCTURE OF $\alpha ^{\prime }-$NaV$_2$O$%
_{\text{5 }}$ COMPOUND: 1D OR 2D?}
\date{3-rd September, 1999.}
\maketitle

\begin{abstract}
The AV$_n$O$_{2n+1}$ mixed valence compounds ($n=1,$ $A=Na$) are classified
as dimerized layered system with strongly interacting $d-$electrons of
vanadium ions. The derived band gaps, energy dispersion relations and
density of electronic states are in a good agreement with available
experimental and theoretical data. The correlated band gap provides the
insulating state of the high-temperature $\alpha ^{\prime }-$NaV$_2$O$_5$
phase whereas the state, earlier misrepresented as the spin-Peierls state,
is governed, in fact, by opening of the Coulomb gap.
\end{abstract}

The spin-Peierls phase transition was first observed in one-dimensional
organic salts [1]. Since its discovery in quasi-1D material \textit{CuGeO}$%
_3 $ [2, 3], there were a lot of efforts to find out a spin-Peierls behavior
in other inorganic materials. The \textit{AV}$_n$\textit{O}$_{2n+1}$ family (%
$A$ stands for alkali or alkali earth element ) has been quite perspective
in that respect. In the best studied $\alpha ^{^{\prime }}-$phase of \textit{%
NaV}$_2$\textit{O}$_5,$ the opening of a spin gap $\Delta _0\sim 80-110K$
was observed at $T_c\approx 34-36K$ $\left[ 4-6\right] $. But at present it
has become evident, that the spin-Peierls scenario can not adequately
describe the properties of $\alpha ^{^{\prime }}-NaV_2O_5$ oxide. A summary
of controversies is given in Ref.$\left[ 7\right] $: the enlarged entropy of
the transition, an enhanced value $2\Delta _0/T_c$, weak dependence of $T_c$
on magnetic field, significant increase of the thermal conductivity below $%
T_c$ in contrast with the conventional spin-Peierls transition.

The $\alpha ^{^{\prime }}-$ $NaV_2O_5$ crystal with a stable $P_{mmn}$($%
D_{2h}^{13}$) symmetry contains the quarter-filled dimers $%
V^{4+}(3d^1)-V^{5+}(3d^0)$ ($T<T_c$ ) or $V^{4,5+}-V^{4,5+}$ ($T>T_c$),
forming rungs of two-leg ladders in the $\left( ab\right) $ plane (Fig.1).
Pyramides $VO_5-VO_5$ are arranged in alternating layers made by vanadium
ions and basal oxygens separated by layers formed by $Na$ and apical
oxygens. For lattices compressed along the $c-$axis ($\alpha ^{^{\prime
}}-NaV_2O_5$ is indeed the case) the crystal field lifts the $t_{2g}-$%
degeneracy of the vanadium 3$d$-levels with the lowest energy of the single $%
d_{xy}$-orbital. Theoretical models usually used for $\alpha ^{^{\prime
}}-NaV_2O_5$ [8-11] are based on fascinating features of a quasi-1D spin
ladder picture beyond the discussion about the electronic structure. Local
density analyses [12, 13] have shown that eigenstates near the Fermi energy
are constructed mainly from the $d_{xy\text{ }}$orbitals of $V$ atoms and
revealed the split character of the 3d-band with pronounced peaks in the
density of electronic states. But the computations could not establish the
character of the peaks in the band structure: do they have $\mathit{1D}$, $%
\mathit{2D}$ or $\mathit{3D}$ character? According to them the Fermi level
lies inside the conducting band providing the metallic phase, which is in a
disagreement with experimental observations.

In this letter we analyze the electronic structure of $\alpha ^{^{\prime
}}-NaV_2O_5$. The phase transition at $T_c$ is shown to be not a pure
spin-Peierls type but rather it is connected with the opening of the Coulomb
gap in the electronic spectrum. The character of the high-temperature
insulating phase has been identified via strong electron interactions. Our
approach is based on the hypothesis that the $\alpha ^{^{\prime }}-NaV_2O_{5%
\text{ }}$properties are governed by the electron-electron correlations $U$, 
$t_a$ (intrarung/intradimer electron hopping integral), $t_b$ (the amplitude
of an electron hopping along legs in crystallographic $b$-direction), $t_{d%
\text{ }}$(the hopping along ladder diagonals), $t_{xy\text{ }}$ (interdimer
hoppings between vanadium ions on nearest ladders) (Fig.1).

Nonspherical angular part $d_{xy\text{ }}$of the $3d-$electron
wave-functions $\Psi (\overrightarrow{r})$ provides the hopping anisotropy
and a layered structure of $\alpha ^{^{\prime }}-NaV_2O_{5\text{ }}$ with
crystallographic $\left( ab\right) $ planes. The small parameter $r_B/a$ ($a$
is a lattice constant and $r_B$ is a vanadium ionic radii) enables us to
evaluate the direct $V-V$ hopping integral $t_{xy}$ for $d-$ electrons with $%
\Psi (\overrightarrow{r})=f\left( r\right) d_{xy}$. The Gaussian radial part 
$f\left( r\right) $ allows to calculate hopping integrals with any required
accuracy as power series of $\left( r_B/a\right) ^2$ [14]. The estimates
show the strong influence of the $V^{4+/5+}$ ion core on an electron
hopping. Therefore we will distinguish the $t_{xy}$ magnitudes especially in
ordered valence state (Fig.1) as $t_{xy}^{bc}>t_{xy}^{bm}>t_{xy}^{qm}$. The
enhanced values of indirect $t_{a,\text{ }b,\text{ }d}$ hoppings are
influenced by intermediate oxygens. Since the present evaluations of the
indirect hopping $t_b$ lead to a substantial ambiguity [8, 10, 12, 13], the
recent infrared reflectance studies of $\alpha ^{^{\prime }}-NaV_{\text{2}%
}O_{\text{5 }}$[15, 16] are more suitable to extract the intradimer hopping
amplitude as $t_a\simeq 0.35eV.$ The order of magnitude estimate of the
other electron hoppings gives $t_b\sim t_d\sim t_a/2$. As for on-site $%
d_{xy} $-$d_{xy}$ interactions, they are taken to be infinite (the value of
the Anderson-Hubbard on-site parameter is believed to be $U=4-6$ $eV$ ) and
somewhat weaker Coulomb interactions between neighbors simply shift on-site
one particle energies in the charge ordered phase.

In such a way the present approach is based on the energy scale $U>>t_a$%
\TEXTsymbol{>}$t_b$\TEXTsymbol{>}$t_d$\TEXTsymbol{>}$t_{xy}$ and we classify
the $\alpha ^{^{\prime }}-NaV_2O_{5\text{ }}$oxide as a strongly correlated
dimerized electron system. For vanadium $d_{xy}$-electrons one can carry out
the simplest fermion mapping to projected X-operators $\left[ 17,18\right] $
describing intravanadium transitions between the one-particle ground and an
empty polar states. Applying the X-operator machinery $\left[ 19\right] $,
one can derive the tight-binding energy bands for strongly correlated
carriers. Neglecting scattering effects, calculations have been done in the
first order of the perturbative hopping energy, namely the correlated energy
bands $\xi (p)$ have been extracted from zeros of the inverse Green's
function $D^{-1}(i\omega ,p)=D^{\left( 0\right) -1}(i\omega )+%
\overleftrightarrow{t}(p)$ ($i\omega \rightarrow \xi +i\delta $), where $%
\overleftrightarrow{t}(p)$ is the corresponding matrix of hopping integrals.
The arrangement of dimers in $\alpha ^{^{\prime }}-NaV_2O_{5\text{ }}$is
closer to a triangular lattice. So, our main strategy is based on the
assumption that the $V_2$- rungs form an ideal triangular lattice. \textit{%
Below} $T_c$ the $\alpha ^{\prime }-NaV_2O_5$ is in an \textit{ordered
valence phase} whereas \textit{above} $T_c$ it is in a \textit{mixed valence
state}.

\textit{At }$T<$\textit{\ }$T_c$ vanadiums are packed in eight sublattices $%
a,b,c,d,m,n,p,q$ (Fig.1). In a zigzag charge/spin order $d_{xy}-$electrons
acquire on-site energy shifts $\varepsilon _{a,d,q,m}=-\varepsilon
_{b,c,n,p}\equiv -\varepsilon $, influenced by neighboring Coulomb repulsion 
$V$: $\varepsilon =V\Delta n$ ($\Delta n$ is the charge disproportionation
on nearest neighbors $V^{4+\Delta n\text{ }/\text{ }5-\Delta n}$). At
positions $a,b,m,n$ and $p,q,c,d$ the $d_{xy}$-electrons have spin
projections $down$ and $up$, respectively. This situation, in parallel with $%
U=\infty $, allows to consider spinless electrons. The resulting energy
dispersions are plotted in Fig.2. The flatness of an antibonding band is
caused by the diagonal intraladder hopping $t_d$. Four electrons from the $%
V^{4+}-V^{5+}$ dimers of a monoclinic unit cell occupy bonding branches
completely. The insulating charge (Coulomb) gap $\Delta _C$ is provided by
the zigzag ordered one-particle energies, $\mp \varepsilon $ , hopping
parameters $t_{a,xy}$ and competing interdimer hoppings $t_{b,d}$. For
realistic magnitudes of $t_{a,b,d,xy}$ and $\varepsilon $ in $\alpha
^{\prime }-NaV_2O_5$ one can conclude that in low-temperature phase the $%
\Delta _C=1eV$ (Fig.2). Of special interest is the role of unequal
interladder hoppings $t_{xy}$: for $t_{xy}^{bc}=t_{xy}^{bm}+\delta $ and $%
t_{xy}^{qm}=t_{xy}^{bm}-\delta $ we have revealed the nonmonotonic $\Delta
_C $ dependence on parameter $\delta $. For $t_{xy}\neq 0$ a finite energy
threshold, $\varepsilon _c$, is necessary to produce the zig-zag order. The
critical Coulomb magnitude $V_c\sim 0.2$ $eV$ given in Ref.$\left[ 11\right] 
$, corresponds to our $\varepsilon _c=28.56meV$ for $t_{xy}=0.06eV$, $\delta
=0.01eV$. Note worthy that for these parameters the Coulomb gap value
coincides with the critical temperature of the so called ''spin-Peierls''
transition: $\Delta _C\left( \varepsilon _c\right) =35K$. Neglecting
interladder hoppings the gap in an electronic spectrum can be approximated
explicitly as 
\begin{equation}
\Delta _C=\sqrt{\varepsilon ^2+\left( t_a+2t_b\right) ^2}+\sqrt{\varepsilon
^2+\left( t_a-2t_b\right) ^2}-4t_d\text{.}  \tag{1}
\end{equation}
The Eq.(1) is the extension of the splitting in terms of the
''charged-magnon'' scenario for a single $V^{4+}-V^{5+}$ rung ($t_{b,d}=0$)
used in Refs.$\left[ 10,16\right] $. We have established also, that an often
discussed chain-type order for $\alpha ^{\prime }-NaV_2O_5$, $\varepsilon
_{a,p,c,m}=-\varepsilon _{b,q,n,d}\equiv -\varepsilon $ ($c.f.$, Fig.1),
does not cause the $\Delta _C$ formation below $T_c$, to trigger the phase
transition of interest.

\textit{At }$T_{\text{ }}>T_{c\text{ }}$a quarter-filled highly dimerized
compound $\alpha ^{\prime }-NaV_2^{4.5+}O_5$ can be described by a
half-filled Hubbard like model for bonding electrons with two dimers/sites
in a orthorhombic unit cell (Fig.1). For $U=\infty $ an effective
Anderson-Hubbard parameter of this model is simply the gain of the
intradimer kinetic energy: $2t_a$. Then the energy bands are formed by the
four one-particle branches 
\begin{eqnarray}
\frac{\xi _p^{+}}{t_b+t_d} &=&\varepsilon _p^{\pm }+\sqrt{\left( \frac{t_a}{%
t_b+t_d}\right) ^2+\left( \varepsilon _p^{\pm }\right) ^2},\text{ }\frac{\xi
_p^{-}}{t_b+t_d}=\varepsilon _p^{\pm }-\sqrt{\left( \frac{t_a}{t_b+t_d}%
\right) ^2+\left( \varepsilon _p^{\pm }\right) ^2},  \tag{2} \\
&&  \nonumber
\end{eqnarray}
where dimensionless tight-binding non-correlated energies are

\begin{equation}
\varepsilon _p^{\pm }=-\cos p_y\pm 2t\cos \frac{p_y}2\cos \frac{p_x\sqrt{3}}2%
\text{ }\left( t=\frac{t_{xy}}{2(t_b+t_d)}\right) \text{.}  \tag{3}
\end{equation}
So, the standard energies (3) are split by the electron interactions. The
lower, $\xi _p^{-}$, and the upper, $\xi _p^{+}$, bands are split due to the
presence of a pair of $V^{4.5+}-V^{4.5+}$ dimers/sites in a high-temperature
unit cell. The two $d_{xy}-$electrons from a unit cell occupy the lower pair 
$\xi _p^{-}$ of the correlated energy bands.

The non-correlated energy dispersions, Eq.$\left( 3\right) $, reflect the
main peculiarities of the reported ''spaghetti'' pictures $\left[
12,13\right] $ rather well. They are in the ranges $-1-2t\leq \varepsilon
_p^{-}\leq 1$ and $-1\leq \varepsilon _p^{+}\leq 1/2+t.$ The upper bonding
branch, $\varepsilon _p^{+}$, has a hyperbolic metric in a proximity of the $%
\Gamma \left( 0,0\right) $-point of the Brillouine zone. The electrons with
these energies have a 2D character of motion. The electrons with bonding
energies, $\varepsilon _p^{-}$, posses 1D features. If it were metallic
carriers, the $\varepsilon _p^{-}$ and $\varepsilon _p^{+}$ would have
provided the quasi-2D saddle and the quasi-1D saddleless portions of the
Fermi surface, respectively.

Above $T_c$ the electron density of states is positioned in the ranges of
correlated energies $\xi _p^\alpha $ ($\alpha =\pm $). For dimensionless
energies, $\omega _\alpha =$ $\xi _p^\alpha /(t_b+t_d)$ (Eqs.$\left(
2\right) $), it is given analytically by equations

\begin{eqnarray}
&&\rho ^\alpha \left( -1-2t-S+T_\alpha \leq \omega _\alpha \leq
-1+2t-S+P_\alpha \right) =  \nonumber \\
&&  \tag{4}
\end{eqnarray}

\[
=\frac 1{\pi ^2\sqrt{k_{_\alpha }t_{}}}[1+\frac{q_\alpha ^2}{4\left( \alpha
\left| \omega \right| +S\right) ^2}]K(q_\alpha ); 
\]

\[
\rho ^{_\alpha }(-1+2t-S+P_\alpha \leq \omega _\alpha \leq 1+S+Q_\alpha )= 
\]

\begin{eqnarray}
&=&\frac 1{\pi ^2q_{_\alpha }\sqrt{k_\alpha t_{}}}[1+\frac{q_\alpha ^2}{%
4\left( \alpha \left| \omega \right| +S\right) ^2}][K(\frac 1{q_\alpha
})\vartheta (1-S+Q_\alpha -\omega _\alpha )+  \nonumber \\
&&+F(\arcsin \frac 1{a_{_\alpha }};\frac 1{q_{_\alpha }})\vartheta (\frac
12+t-S+R_\alpha )],  \tag{5}
\end{eqnarray}
where $\vartheta $ is a Heaviside step function, $S=\sqrt{\tau ^2+\left(
1+2t\right) ^2}-\sqrt{\tau ^2+1}-t$, $T_\alpha =\alpha \sqrt{\tau ^2+\left(
1+2t\right) ^2}$, $P_\alpha =\alpha \sqrt{\tau ^2+\left( 1-2t\right) ^2}$, $%
Q_\alpha =\alpha \sqrt{\tau ^2+1}$, $R_\alpha =\alpha \sqrt{\tau ^2+\left(
1/2+t\right) ^2}$, ($\tau \equiv t_a/(t_b+t_d)$). In Eqs.$\left( 4,5\right) $
the elliptic integrals $F$ and $K$ have modulus $q_\alpha =\sqrt{[2t\left(
t+k_\alpha \right) +1-(\omega _0^\alpha )^2]/k_\alpha t}/2$ and $\arg $ument 
$a_\alpha =\sqrt{k_\alpha \left( 1+\omega _0^\alpha \right) \left(
t+k_\alpha \right) /[2t\left( t+k_\alpha \right) +1-(\omega _0^\alpha )^2]}$
with $k_\alpha =\sqrt{2\left( 1-\omega _0^\alpha \right) +t_{}^2}$, $\omega
_0^\alpha =[\left( \alpha \left| \omega \right| +S\right) ^2-\tau
^2]/[2\left( \alpha \left| \omega \right| +S\right) ]$. The Eqs.(4, 5) are
valid if the $\varepsilon _p^{+}$-band, Eq.$\left( 3\right) $, is inside the
energy interval of the $\varepsilon _p^{-}$-band, $i.e.$ at the hopping
parameters obeying the realistic for $\alpha ^{^{\prime }}-$NaV$_2$O$_5$
constraint $t_{xy}<t_b^{}+t_d$. The main panel of Fig.3 displays the density
of correlated electron states with a gapped electronic structure at $T>$ $%
T_c $. In a limiting non-correlated case, the density of states (inset)
reproduces the essentials of the first principle computations [12]. The
overlap of the energy ranges for the electronic dispersions, Eqs.(2, 3),
leads in Fig.3 to peculiarities of the electronic structure at $%
L_3=1/2+t+R_{-}-S$, $U_3=1/2+t+R_{+}-S$ in the main panel and at $1/2+t$ in
inset. \textit{Logarithmic divergencies} inside the band at $\varepsilon
=-1+2t$ and $\varepsilon =1$ (inset), and at $L_2=-1+2t+P_{-}-S$ and $%
U_2=-1+2t+P_{+}-S$ (main panel) \textit{are clear manifestations of the 2D
electronic structure of }$\alpha ^{\prime }-$\textit{NaV}$_2$\textit{O}$_5$%
\textit{\ compound}. We would like to emphasize that \textit{in the
one-dimensional limit }($t_{xy}\rightarrow 0$) the electron density of
states is taking features of a single spin-ladder without any logarithmic
peaks, \textit{the divergencies are becoming square-root like} with
positions at the band edges. Under the strong electron interactions the
lower correlated bands, Eqs.(2), are completely occupied by the electron
pair from an orthorhombic unit cell (Fig.1). Finally it results in the
appearance of the correlated band gap (Fig.3): 
\begin{eqnarray}
\Delta _g &=&\min \xi _p^{+}\left( \varepsilon _p^{-}\right) -\max \xi
_p^{-}\left( \varepsilon _p^{+}\right)  \nonumber \\
&=&\sqrt{t_a^2+(t_b+t_d)^2}+\sqrt{t_a^2+\left( t_b+t_d+t_{xy}\right) ^2}%
-2\left( t_b+t_d\right) -t_{xy}.  \tag{6} \\
&&  \nonumber
\end{eqnarray}
Substituting to Eq.(6) the realistic values of hopping parameters, $%
t_a=0.35eV$, $t_b=0.15eV$, $t_d=0.1eV$, $t_{xy}=0.06eV$, one can derive the
numerical value of the high-temperature gap in $\alpha ^{\prime }-NaV_2O_5$
as $\Delta _g=0.34eV$.

In summary, the $\alpha ^{\prime }-NaV_2O_5$ band structure has been
analyzed. The analysis of derived $\xi _{\overrightarrow{p}}^{\pm }$ curves
and density of electronic states in an explicit form leads to the conclusion
about pronounced 2D features. At $T_{\text{ }}<T_{c\text{ }}$the zigzag
order redistributes large $V^{4+}$ and small $V^{5+}$ ions and it is
accompanied by the Coulomb gap $\Delta _C$ (see Fig.2 and Eq.$\left(
1\right) $). Its estimated magnitude $\Delta _C\simeq 1eV$ (for $\Delta
n=0.8 $, $V=0.8eV$ $\left[ 20\right] $, $t_a=0.35eV$, $t_b=0.15eV$, $%
t_d=0.1eV$) corresponds to the observed strong absorption of the light $%
\left[ 17\right] $. At $T_{\text{ }}>T_{c\text{ }}$the correlated band gap $%
\left( 6\right) $ provides an insulating state of $\alpha ^{\prime }-NaV_2O_5
$. The studies, reported in Ref.$\left[ 21\right] $, give an experimental
evidence of the clear semiconducting behaviour of $\alpha ^{\prime }-NaV_2O_5
$ below and above $T_c$ with the increased dimensionality of an electron
transport.

In $\alpha ^{\prime }-NaV_2O_5$ the strong interplay between charge, spin
and lattice degrees of freedom should be forthcoming $\left[ 22\right] $.
The electron-lattice interactions renormalize an energy scale. Starting from
ideas about on-site $\left[ 23\right] $ and inter-site $\left[ 24\right] $
electron pairs it is possible to show that the $V^{4+}-V^{4+}$ dimers are
the shortest $\left( S\right) $, whereas $V^{5+}-V^{5+}$ones are the longest
ones $\left( L\right) $ ($c.f.,$Fig.1). This conclusion is also consistent
with the X-ray and neutron diffraction data $\left[ 25\right] $ indicating
the presence of the low-temperature modulated sequence $S-L-L-L-S-L...$ in $%
\alpha ^{\prime }-NaV_2O_5$.

According to Eqs.$\left( 3\right) $ there are two periodicities along $b$%
-axis in agreement with ARPES\ data $\left[ 26\right] $. At $T<$ $T_c$ the
charge order has been obtained without invoking the exchange $J$-terms.
However, as a consequence of charge order the alternative exchange
antiferromagnetic interactions open the spin-gap $\left[ 7\right] $.
Diagonal hopping parameters $t_d,$ $t_{xy}^{bc},t_{xy}^{bm},t_{xy}^{qm}$
cause exchange interactions, responsible for splitting of magnon modes
observed in inelastic neutron scattering $\left[ 27-29\right] $. If $J$
constants are much smaller than electron hopping, the spin-dependent terms
can be treated as a perturbation resulting in a spin-charge separation for
1D Hubbard or $t-J$ models $\left[ 30\right] $. From that point of view it
is interesting to consider an interpretation of a strong temperature induced
modification of the spectral intensity seen by the ARPES in $NaV_2O_{5\text{ 
}}\left[ 26,31\right] $. In terms of the 1D $t-J$ model the authors $\left[
31\right] $ described the experiment as an evidence of availability of
spinon and holon Fermi surfaces. Our study of electronic structure has
revealed an importance of interladder couplings in $\alpha ^{\prime
}-NaV_2O_5$.

Let us also note that the $A^{+}V_{\text{2}}O_{\text{5 }}$-family can be
treated as an electron counterpart of striped layered cuprates. An angular
symmetry of $d_{x^2-y^2}$-wave function of copper holes in layered high-T$_c$
cuprates has a similarity with the $d_{xy}$ - wave function of vanadium
electrons. However, in vanadates the role of $d-p$ hybridization is
diminished but the role of electron correlations is enhanced in contrast to
cuprates.

In conclusion, the authors would like to thank Yvan Bruynseraede for
stimulating discussions. This work is supported by the Belgian IUAP and
Flemish GOA and FWO Programs. 
\[
\]

\textbf{On leave from :}

$^1$\textit{N. S. Kurnakov Institute of the General and Inorganic Chemistry
of the}

\textit{\ Russian Academy of Sciences, Leninskii prospect 31, 117 907 Moscow,%
}

\textit{\ RUSSIA}

$^2$\textit{Institute of Physics, 11080 Belgrade, P.O. Box 68, YUGOSLAVIA}

$^3$\textit{Institute of Engineering Physics, Hanoi University of
Technology, 10 000 Hanoi, VIETNAM}\newpage\ 

\begin{center}
\textbf{References}
\end{center}

$\left[ 1\right] $ T. Ishiguro, K. Yamaji and G. Saito ,\textsl{Organic
Superconductors }(Springer Series in Solid-State Sciences, v.88, Springer,
1998) and Refs. there.

$\left[ 2\right] $ G. Petrakovskii $et$ $al.,$ JETP \textbf{71,} 772 (1990)
- we note that their analysis had been done in terms of the
antiferromagnetic phase transition.

$\left[ 3\right] $ M. Hase, I. Terasaki, and K. Uchinokura, Phys. Rev. Lett.%
\textsl{\ }\textbf{70, }3651\textbf{\ } (1993).

$\left[ 4\right] $ M. Isobe and Y. Ueda,\textsl{\ }J. Phys. Soc. Jpn. 
\textbf{65, }1178 (1996).

$\left[ 5\right] $ M. Weiden $et$ $al.,$ Z. Phys. B \textbf{103, }1 $\left(
1997\right) $.

$\left[ 6\right] $ T. Ohama $et$ $al.,$J. Phys. Soc. Jpn.\textsl{\ }\textbf{%
66, }3008 (1997); Phys. Rev.B \textbf{59,} 3299 (1999).

$\left[ 7\right] $ M.V. Mostovoy and D.I. Khomskii, cond-mat/9806215 $\left(
1998\right) $.

$\left[ 8\right] $ P. Horsch and F. Mack,\textsl{\ }Eur. Phys. J. B \textbf{%
5,} 367 (1998).

$\left[ 9\right] $ P. Thalmeier and P. Fulde, cond-mat/9805230 $\left(
1998\right) $.

$\left[ 10\right] $ S. Nishimoto and Y. Ohta, J. Phys. Soc. Jpn. \textbf{67, 
}3679 (1998).

$\left[ 11\right] $ H. Seo and H. Fukuyama, J. Phys. Soc. Jpn., \textbf{67, }%
2602 (1998).

$\left[ 12\right] $ H. Smolinskii $et$ $al.,$\textsl{\ }Phys. Rev. Lett%
\textsl{.} \textbf{80,} 5164 (1998).

$\left[ 13\right] $ Z.S. Popovic and F.R. Vukailovic, Phys. Rev. B \textbf{%
59, }5333 (1999).

$\left[ 14\right] $ V.A. Ivanov, J. Phys.: Condens. Matter \textbf{6, }2065
(1994); Physica C \textbf{185-189,} 1635 (1991).

$\left[ 15\right] $ D. Smirnov \textit{et al.}, Physica B \textbf{259-261,}
992 $\left( 1999\right) $.

$\left[ 16\right] $ A. Damascelli $et$ $al.$, cond-mat/9906042 $\left(
1999\right) $; A. Damascelli $et$ $al.$, Phys. Rev. Lett\textsl{.} \textbf{%
81,} 918 (1998).

$\left[ 17\right] $ S. Okubo,\textsl{\ }Prog. Theor. Phys., \textbf{27, }949
(1962).

$\left[ 18\right] $ J. Hubbard, Proc Roy. Soc. A \textbf{276,} 238 (1963), $%
ibid.$ \textbf{277,} 237 (1964).

$\left[ 19\right] $ V.A. Ivanov,\textit{\ }Physica B\ \textbf{186-188,} 921
(1993); Physica C \textbf{271, }127 (1996); in \textit{Studies of High
Temperature Superconductors,} edited by A. Narlikar, Vol. \textbf{11} (Nova
Science Publishers, New York, 1993 ), p.331.

$\left[ 20\right] $ M. Cuoco, P. Horsch, and F. Mack, cond-mat/9906169
(1999). \textit{\ }

$\left[ 21\right] $ J. Hemberger \textit{et al.} Europhysics Lett. \textbf{42%
}, 661 (1998) ; M. Lohmann $et$ $al.$ Physica B\ \textbf{259-261}, 983
(1999).

$\left[ 22\right] $ J. Riera and D. Poilblanc, Phys. Rev. B \textbf{59,}
2667 $\left( 1999\right) $.

$\left[ 23\right] $ P.W. Anderson, Phys. Rev. Lett. \textbf{34,} 953 $\left(
1975\right) $.

$\left[ 24\right] $ B.K. Chakraverty $et$ $al.$, Phys. Rev. B \textbf{17,}
3781 $\left( 1978\right) $.

$\left[ 25\right] $ T. Chatterji $et$ $al.,$ Solid State Commun. \textbf{108,%
} 23 $\left( 1999\right) $.

$\left[ 26\right] $ K. Kobayashi $et$ $al.,$ Phys. Rev. Lett. \textbf{80,}
3121 $\left( 1998\right) $.

$\left[ 27\right] $ T. Yoshihama $et$ $al.$, J. Phys. Soc. Jpn. \textbf{67,}
744 (1998).

$\left[ 28\right] $ C. Gros and R. Valenti, Phys. Rev. Lett. \textbf{82},
976 $\left( 1999\right) $.

$\left[ 29\right] $ P. Thalmeier and A.N. Yareshko, cond-mat/9904443 $\left(
1999\right) $.

$\left[ 30\right] $ S. Sorella and A. Parola, J. Phys.: Condens. Matter 
\textbf{4}, 3589 (1992); H. Suzuura and N. Nagaosa, Phys. Rev. B \textbf{56}%
, 3548 $\left( 1997\right) $.

$\left[ 31\right] $ K. Kobayashi $et$ $al.,$ Phys. Rev. Lett. \textbf{82,}
803 $\left( 1999\right) $.\newpage\ 

\begin{center}
\textbf{Figure captions}
\end{center}

\textbf{Fig.1.}

The schematic view of $\alpha ^{\prime }-NaV_2O_{5\text{ }}$. Each
dimer/rung is replaced by a circle. The inter(intra)dimer hopping $t_{b\text{
}}$($t_a$) in the $b$($a$)-direction is set along the $y\left( x\right) $%
-axis. The distances at room temperature between the nearest $V$-ions on
neighboring dimers/rungs are $3.04\AA $\ and the leg constant is $3.61\AA $.
The dimer size is $3.44\AA $. Oxygen $p-$wave functions (opened) enhance the
hopping $t_{d\text{ }}$along ladder diagonals. For $T>T_c$ : the
orthorhombic unit cell with two dimers is shown in lower panel. For $T<T_c$:
the size of arrows (lower panel) reflects the charge disproportionation $%
\Delta n=n_{a,d,m,q}-n_{b,c,n,p}$ in the monoclinic unit cell; the shaded
portions have a zigzag order.

\textbf{Fig.2.}

The tight-binding energy dispersions for correlated $d_{xy}$-electrons in $%
\alpha ^{\prime }-NaV_2O_5$ \textit{below }$T_c$ for parameters $t_a=0.35eV,$
$t_b=0.15eV,$ $t_d=0.1eV,$ $t_{xy}^{bm}=0.06eV,\delta =0.01eV$ and $%
\varepsilon =V\Delta n$ $(V=0.8eV,\Delta n=0.8)$. Momenta are given in units 
$\left| p_x\sqrt{3}\right| =\left| p_y\right| =\pi $ of the Brillouine zone
boundaries, the Fermi energy, $E_F=0$, is inside the Coulomb gap $\Delta
_C=1 $ $eV$.

\textbf{Fig.3.}

The high-temperature $\left( T>T_c\right) $ electron density of states as a
function of dimensionless energies $\xi $ $/(t_b+t_d)$, $E_F=0$ (main
panel). The inset shows the density of states for non-interacting bonding
electrons. The Latin letters denote the energies: $L_1=-1-2t-S+T_{-},$ $%
L_2=-1+2t-S+P_{-},$ $L_3=1/2+t-S+R_{-},$ $L_4=1-S+Q_{-},$ $%
U_1=-1-2t-S+T_{+}, $ $U_2=-1+2t-S+P_{+},$ $U_3=1/2+t-S+R_{+},$ $U_4=1-S+Q_{+}
$, where parameters $S,$ $T_{\pm },$ $P_{\pm },$ $Q_{\pm },$ $R_{\pm }$ are
done in Eq.$\left( 5\right) $ and $t=t_{xy}/[2(t_b+t_d)]$.

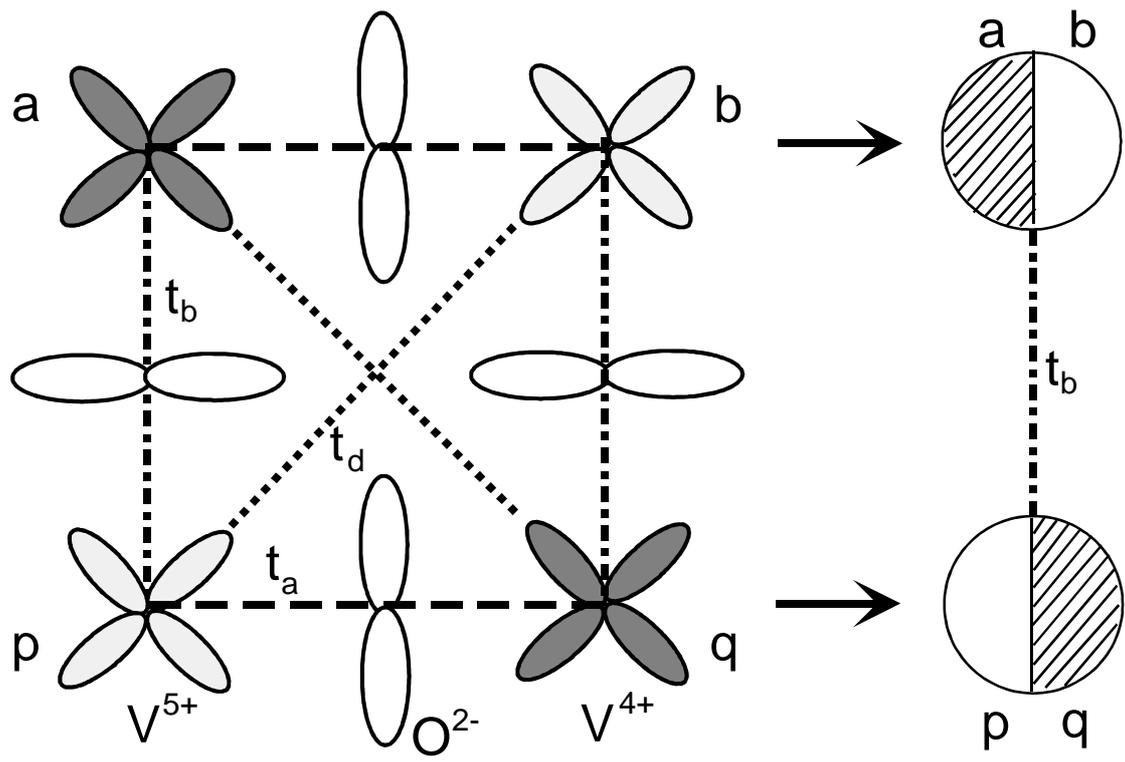
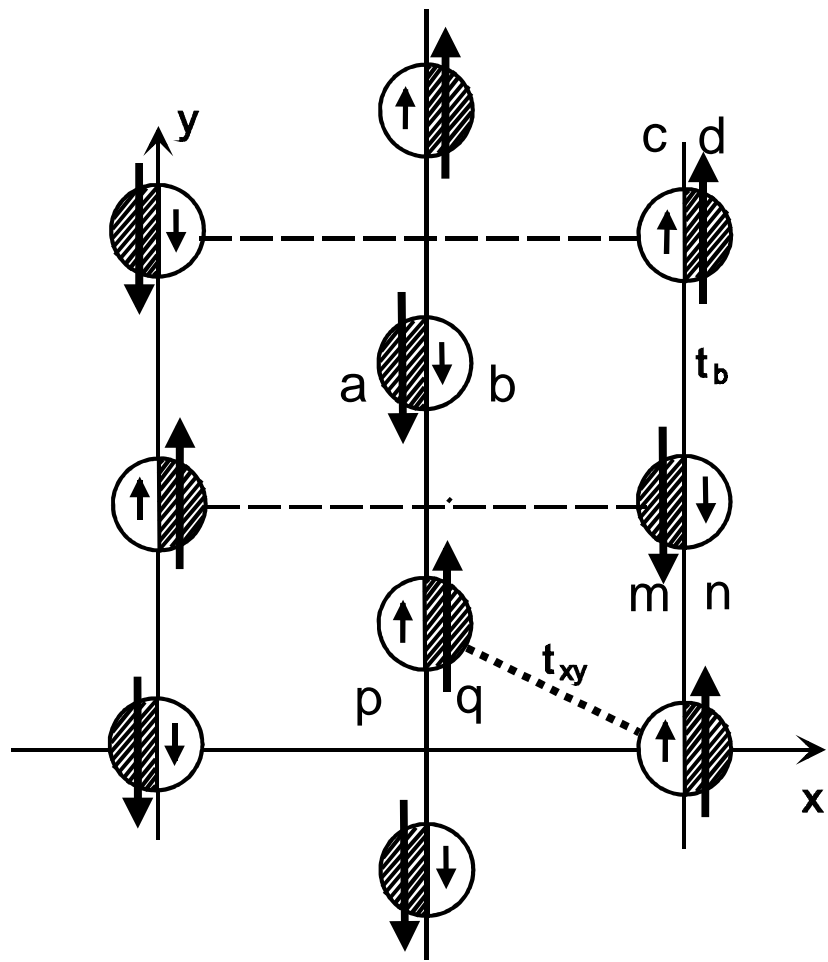
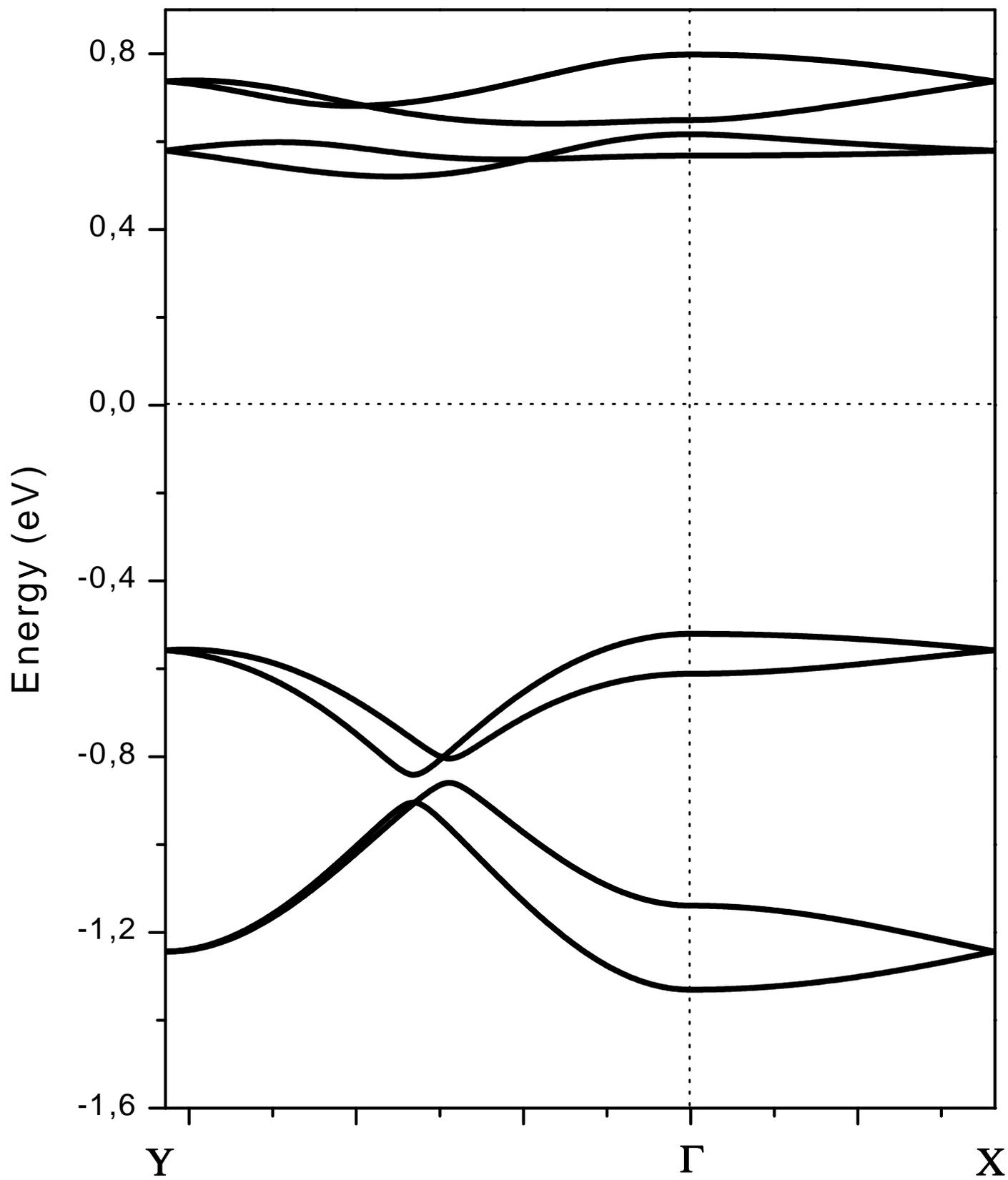
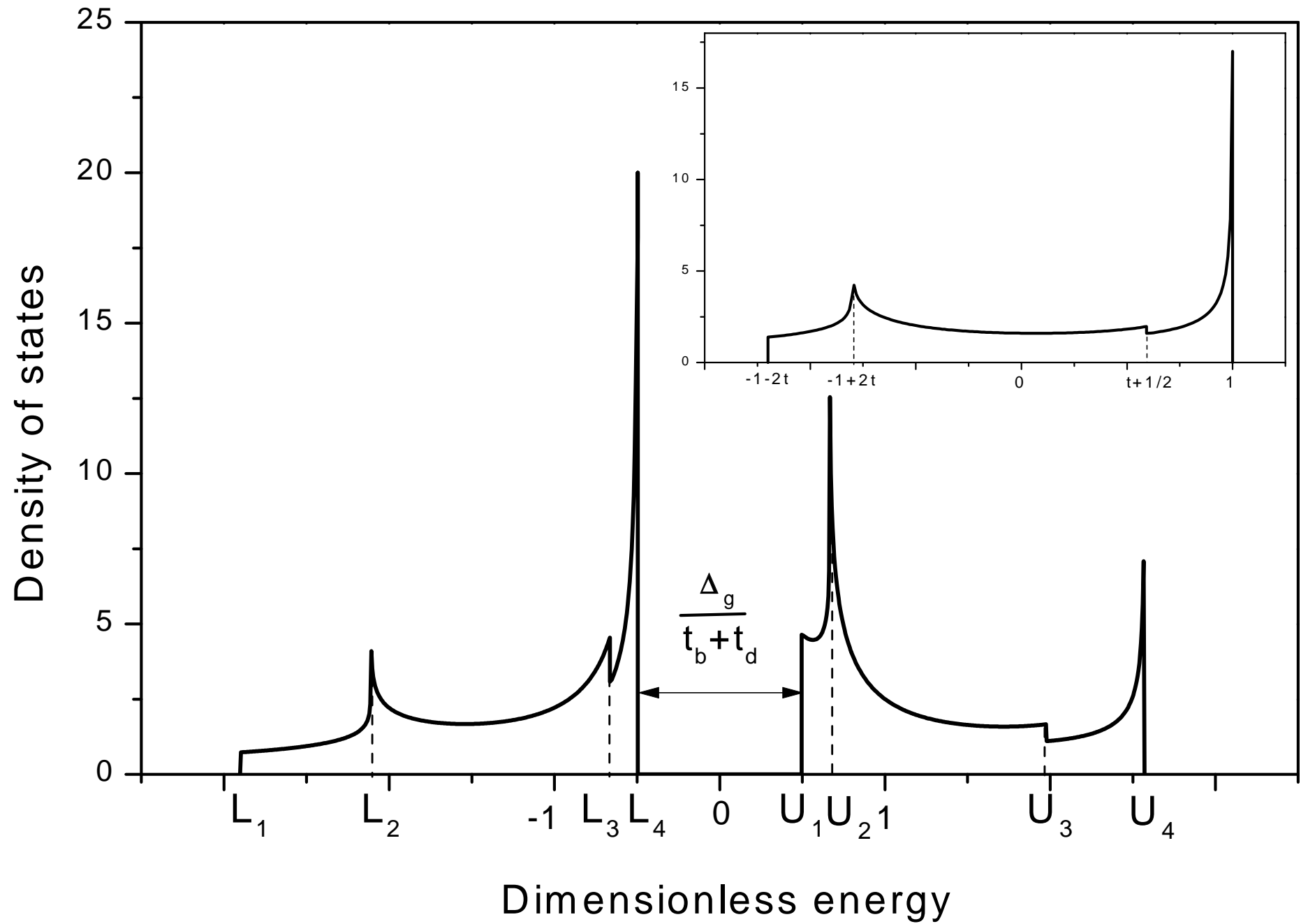
\end{document}